# Exploring the interplay between population profile and optimal routes in U.S. cities

# HUMAN BEHAVIOUR IN CITY DEVELOPMENT

**Diego Ortega**[1*] **& Elka Korutcheva**[1,2]

[1] **Fundamental Physics Department, National University of Distance Education (UNED), Madrid, 28040, Spain**

[2] **G. Nadjakov Institute of Solid State Physics, Bulgarian Academy of Sciences, Sofia, 1784, Bulgaria**

*email: blackgats@gmail.com , dortega144@alumno.uned.es

## ABSTRACT

Cities have developed over time alongside advancements in civilization, focusing on efficient travel and reducing costs. Many studies have examined the distinctive features of urban road networks, such as their length, efficiency, connection to population density, and other properties. However, the relationship between car routes and population in city structures remains unclear.

In this study, we used the center of mass for each city tract, defined by the US Census, as the origins and destinations for our itineraries. We calculated travel time, and both Euclidean and travel distances for sixty major cities. We discovered that the total sum of all routes adheres to an urban law. The distribution of these car journeys follows Weibull functions, suggesting that the urban center plays a crucial role in optimizing routes across multiple cities. We also developed a simple point pattern model for the population, which aligns with the well-known decreasing exponential density expression.

Our findings show that the interplay between population and path optimization influences city structure through its center. This study offers a new perspective on the fundamental principles that shape urban design.

## Introduction

From ancient times, human beings have organized themselves socially, first in the form of settlements, and then in cities. For the people who live in them, there has always been a trade-off between the benefits, derived from social gatherings, and the cost associated with transport. From a theoretical perspective, this balance is present in the amorphous settlement model[1], the central market theory[2], the core-periphery framework[3] and the Alonso´s model for land rents in the city[4]. The latter model points out that a decrease in the price of transport leads to an expansion in the city area. This theory explains why it is not until the advent of the combustion engine and the construction of roads that cities take on their current shape[5].

The urban road network has been characterized from different perspectives. If one relates its total length $L$, to the population $P$, for several cities, one finds an exponential expression $L \propto P^{\gamma}$, which defines an *urban law*[1,6]. In general, these expressions establish connections between population size and various urban outcomes such as infrastructures[6], innovation[7], firearm violence[8] or waste generation[9], etc. For urban roads, $\gamma < 1$, which corresponds to the sublinear regime and is due to the fractal nature of the network. The people of the city must be connected; thus, the network must occupy space by filling in the area [6,10]. From the other side, length does not offer information on how well the urban network is built. One of the most used efficiency measures to provide such information is the detour index[11–14]. Given an origin and a destination, it is the ratio between the network and Euclidean distances. For metropolitan areas it is usually related to a distance-decay function, i.e., the longer the trip, the lower the index is[13,14]. On the other hand, contributions from spatial networks theory also point to the fact that urban networks share similarities[15–18]. In this schematic representation of the street network, roads and intersections are substituted by edges and nodes, respectively. The degree distribution of intersection is peaked, and the most interesting information lies in the distribution of betweenness centrality, a measure based on the number of shortest paths that pass through a node[16,18].

Since urban road networks exhibit consistent features irrespective of their location or geographical features, one infers that there are mechanisms influenced by population dynamics. In this scenario, recent models assumed the coevolution of both population and the road network[19–22], thus establishing a central zone and a population density profile which decreases exponentially with distance[19]. Other factors such as the accessibility and housing price[20], the features of the constructed network[21] and the attraction of a central business district[22] have also been studied.

However, none of the previous work focuses on individual mobility patterns. This analysis is usually carried out by cell phone location data[23–26], which are aggregated as an *origin-destination (OD) matrix* $M$[23,27–29]. Such studies are performed by dividing the area into cells of equal size and identifying them by an index. As they usually refer to the flow of people, the element $M_{ij}$ denotes the population moving from cell *i* to cell *j*. This approach allows for an understanding of how a city is structured for its citizens and whether it is monocentric or polycentric[23,24]. Another option is that these matrices encompass the travel times or distances between various city locations[28,29].

Despite all the work done, questions remain open: How does the optimization of travel routes correlate with the factors mentioned earlier? Do most cities share a similar distribution of optimal travel routes? Is this distribution related to the population density profile? Does the city center play a role in the minimization of transport costs?

In this work we propose a simple method to relate the city's population profile to the optimization of travel routes. In the Methods section, we describe this process in detail. For each city we estimate the centers of mass of the census tracts[30], which are areas with approximately 4,000 inhabitants. Then, we calculate the fastest routes connecting all centroids between them. In this way three OD-type matrices are generated corresponding to the travel time, and to the Euclidean and the travel distances. This study is carried out for 60 North American cities exceeding 100,000 inhabitants. We discuss the existence of urban laws for the sum of these matrices, and the existence of a general distribution for optimal routes.

Furthermore, we apply a basic population model to analyze this distribution, revealing that the presence of a population center significantly contributes to the reduction of travel distances.

# Results

In the present study we divide a city into its census tracts and estimate the position of their center of masses. For a pair of centroids, *i* and *j*, we calculate the Euclidean length, $E_{ij}$, and its travel distance for the fastest route, $D_{ij}$, as it is illustrated in Figure 1. Besides, we also estimate the time to move between the two selected points $T_{ij}$. Replicating this process by taking all centroids as origins and destinations, we calculate three OD matrices for each city: **E** and **D** accounting for Euclidean and travel distances, and **T** for the travel time.

**Urban laws**

As we have previously explained, urban laws connect features in cities with the number of urban dwellers. For our case study, we have plotted the total of Euclidean distances between centroids, $\sum_{ij} E_{ij}$, versus the city population. The results are represented in Figure 2a. The correlation is high, $r^2 = 0.95$, and its slope, the scaling exponent of the urban law, is $\gamma_e = 2.43 \pm 0.07$. Therefore, the total Euclidean distance scales super-linearly with the population. This scaling can be explained by simple properties of complete graphs. As the $N$ centroids are linked to each other, the number of edges is given by $N \cdot (N-1)/2$. Since we work with paths, each edge is traversed twice, thus the 2 factor is disregarded. Therefore, the leading term would be $N^2$ and the total route length is proportional to it. From the other side, these nodes are associated with the center of masses of non-overlapping areas, as it is depicted in Figure 1. This imposes a minimum distance between nodes which could increase the supercritical exponent to values even greater than 2.

The procedure applied to the Euclidean lengths is repeated for the travel distances and their corresponding times, i.e. matrices **D** and **T**. This yields the urban laws depicted in Figures 2b and 2c. Both figures show a high correlation coefficient, $r^2 = 0.95$, and similar values for their slopes, $\gamma_d = 2.43 \pm 0.07$ for distances and $\gamma_t = 2.33 \pm 0.07$ for their times. The value of the temporal slope is slightly lower than that of the spatial one. This may be due to our optimization method, which is focused on minimizing journey times rather than travel distances. A notable feature is the concurrence of the supercritical exponents for the Euclidean and travel distances, $\gamma_e = \gamma_d$. This value is linked to number of centroids and their geographical layout, which remains consistent regardless of the considered distance. Consequently, the exponents are equal. However, further exploration of the link between Euclidean and travel distances is possible using the detour index.

**Detour index**

To study the relationship between the total travel distance and the Euclidean one in a generic way, we assumed a relation $\sum_{ij} D_{ij} = C \cdot (\sum_{ij} E_{ij})^\sigma$ between them for the cities. Here, $C$ is the proportionality factor, $\sigma$ is the related exponent and summation runs over all the city centroids labelled as *i* and *j*. This equation is plotted in Figure 3a. The slope value is $\sigma = 0.998 \pm 0.002$ close to one, and the y-intercept is $b = 0.15 \pm 0.01$. Nevertheless, the illustration has logarithmic axes, thus, $\sigma \approx 1$ implies a linear relation between the Euclidean and travel distances. The proportionality constant between both distances is estimated from to the y-intercept value from Figure 3a, *b*, as $C = 10^b$. As it has been previously defined, the detour index for a given route is the ratio of travel length to Euclidean distance. In the present context, we generalize this concept representing the total of travel distances versus the total of Euclidean ones for each city. Hence, as all the set of cities contributes to the calculation of *C*, we define the mean detour index as $\overline{DI} = C = 1.41 \pm 0.09$. The meaning is straightforward: when travelling between two random centroids in any city, the distance over the urban road network is in average 41% greater than the distance travelled in an ideal straight line.

An alternative calculation of the $\overline{DI}$ can be made by averaging over the cities. For each city, all centroids *i* and *j* were taken, and the detour index was estimated for every possible travel between them. The mean of these values was then taken, obtaining an average value for each city. Finally, this process was repeated for all the set of cities. Further calculation details can be found in the methods section. Averaging each city mean, we obtained $\overline{DI} = 1.43 \pm 0.06$, which is close to the previous value. A Gumbel-type curve fits well the samples density histogram as can be seen in Figure 3b. Note that this distribution is typical of extreme values[31] and linked to the roughness field[32]. The detour index is calculated based on the length of the journey on the urban network relative to the distance between the origin and destination in a straight line. Therefore, it is to be expected that the further it departs from the straight line or the more it meanders, the higher the value. The approach allows us to interpret the detour index as a measure of the city's own roughness showing that most of cities belong to the narrow range value [1.35,1.5]. Despite the necessity to adapt to diverse geographical features, including coastal and riverine areas, the proximity of detour index values suggests the existence of a universal pattern in the design of the road network.

**Collapse of Distributions**

To gain a deeper understanding of these patterns, we analyzed the distributions of travel routes that inhabitants follow to optimize their journeys, regardless of the specific city under consideration.

Cities depicted in Figure 4a exhibit a maximum travel distance distribution within the range of [5,20] kilometers, irrespective of their population size. Given that this distance is readily attainable for a vehicle, it can be assumed that the census tracts are situated near one another on average. Furthermore, Figure 4a illustrates that as the population of a given area increases, from Stamford with 135,000 inhabitants to New York with 8,700,000, the maximum density value decreases, widening the graphs. This led us to express the density histogram of the travel distance by taking its mean for each city. In this way, populations are scaled, allowing comparison of profiles between them.

Although some cities show oscillations against the general trend, one can see that there is a tight curve they follow, as it is shown in Figure 4b. The general trend is fitted by a Weibull curve[33]. This distribution models a broad range of random variables, largely in the nature of a time to failure[31]. The fact that the fitted curve is skewed to the right indicates that most values are concentrated in the left region. This is consistent with the hypothesis that journeys on the urban network are designed to be short. It must be highlighted that the aim of our optimal routes is to minimize the travel time, which is usually achieved by keeping the route as short as possible.

The travel time associated to route distances is illustrated in Figure 4c. A comparison of the graph with the one in Figure 4b reveals that the fitted curve is less skewed. It is important to note that a sizable proportion of journeys may be relatively short, but occurs on low-speed roads, which will result in longer journey times. Conversely, a long journey on a motorway may take less time than a shorter distance on a secondary road.

Finally, the Euclidean distance density is shown in Figure 4d, exhibiting a similar behavior to the travel distance depicted in Figure 4b. The slight variation may be attributed to the fact that short journeys are typically conducted via secondary or tertiary roads, and run through residential zones, with a high detour index. On the other hand, longer journeys may occur on motorways, which are straighter, resulting in a distance closer to the Euclidean measure.

It can be concluded that the optimal routes are linked to the relative position of the centroids themselves with respect to each other and to the type of road prevailing between them. Therefore, it is of interest to understand the population distribution in terms of fastest routes.

**Population distribution**

Figure 4d illustrates an average profile for Euclidean distance paths that can be approximated by a Weibull-type curve. The objective now is to develop a simple spatial framework for centroids that can fit this profile. In other words, we want to define a point pattern model whose distances fit the associated Weibull curve. After visually identifying the centroids of the cities, we found a common pattern: cities with a population between 100,000 and 200,000 tend to have one nucleus. To define this center, it is essential to consider two key factors: census tracts do not overlap, and population density decreases radially. More model details can be found in the Methods section.

We have simulated 1,000 cities of 40 centroids, on a $2 \times 2$ units square. The histogram of their Euclidean distances is shown in Figure 5a (Simulated). Then we have fitted a Weibull curve to these points (Simulated fit) and compared it with the Euclidean distances fit from Figure 4d (Euclidean fit). As can be observed, both curves overlap. This suggests that the fundamental mechanisms that shorten distances are the establishment of a minimum area in which a population can settle, modeled as repulsion between centroids, and attraction towards the population center. The relationship between these two factors determines the city fundamental layout, with the structure of the optimized routes accounting for this. For larger populations, the city becomes polycentric although it seeks to replicate the mechanism previously described. In this way, the cost of transport for larger populations can be reduced, while there is a trade-off between attraction to the multiple centers and repulsion due to building limitations.

To understand how the type of city generates the above profile of Euclidean distances, we examine the radial density of centroids for 1,000 runs. The profile follows a decreasing exponential curve, which implies that the city has a high population density in the urban core, as we can derive from Figure 5b and its inset. Consequently, the concentration of population in the city center results in faster trips between locations, reinforcing the relationship between routes optimization and the design of urban networks.

## Discussion

The present study demonstrates how route optimization interrelates with population to shape cities, creating a structure that connects people through short and fast routes.

We have shown that the total sum of optimal routes is related to population with a super-linear scaling of $\sim 2.4$. This contrasts with the exponents relative to total road length typically observed in other studies[6] being $\sim 0.85$. The discrepancy can be attributed to two factors. Firstly, in our calculations, the total number of routes scales with the square of the centroids number. Secondly, a route in path is not equivalent to the road length. To illustrate this, let us consider a road segment between points A and B. This contributes only once to the total road length. However, the total sum of routes calculated may include at least two paths: from A to B and from B to A. Furthermore, this path may also be travelled as part of another journeys, turning out that the same road segment contributes to several routes between different locations.

The average detour index between centroids was calculated by using two methods, yielding values of 1.41 and 1.43. It is particularly useful to compare these results with those obtained in a similar way. Indeed, the detour index has been measured from the centroids of the census tracts to the nearest hospitals[11], giving a value of $\overline{DI} \sim 1.42$. For random points in the 51 most populous metropolitan areas in the Unites States[14] the estimated detour index was $\overline{DI} \sim 1.34$. It should be noted that the detour index decreases with travel distance, given that, in longer journeys, it is possible to find a highway or main road that can straight the route. Thus, the obtained values in our work are characteristic of distances less than 10 km[14]. This highlight that several centroids are in close proximity to each other. Therefore, the proximity between population groups facilitates the minimization of routes. Indeed, the clustering mechanism around a core is so effective that it is adopted by the majority of cities included in our study. Thus, when normalized by their average values, the time and travel distance curves collapse into Weibull functions, as it has been depicted in Figure 4. The functions are right-skewed, indicating that most routes are short in distance and duration. In a recent study, that calculates the detour index considering monocentric and radial structure for cities *a priori*, universal properties of the detour index were discovered[34]. These properties included the collapse of the detour index curves by scaling the journey distance with the city radius. Hence, the importance of the core and periphery model was highlighted once again, explaining why many urban networks have similar characteristics, regardless their geographical location or history. It must be noted that the previously discussed model[34] is monocentric in nature, and cannot be applied to the world's largest megacities, which are characterized by a polycentric layout. In our methodology, both polycentric and monocentric cities collapse over the same curve indicating that regardless of its structure, the objective is to ensure that people are linked with a minimum cost. This leads us to the conclusion that this trait in human behaviour is a pervasive factor in urban design.

Finally, we have developed a simple framework for the distribution of a monocentric population, based on a point pattern for centroids. The proposed model is a hard sphere one with attraction to the city center, which minimizes the fitting error to the adjusted distribution of Euclidean distances in Figure 4d. We have found that the probability of the existence of a centroid depends on the inverse of the distance squared, which can be linked to gravitational models of population[35,36]. Furthermore, the population density profile yielded a decreasing exponential curve, which is one of the expressions that have been used to estimate these distributions since the 1950s[35,37].

In summary, by embracing the route optimization approach, various aspects of cities typically modeled separately were elucidated in a unified manner. The interplay between population and the urban road network emerged as the underlying reason behind the overarching patterns observed in multiple cities.

A direction for future research is the study of European cities. Their evolution is typically based on a center that has acted as a historical nucleus. Another potential area for further investigation could be the development of a point pattern model for polycentric cities that also fits the Euclidean distance distribution presented in this study.

# Methods

### Data sources
Two sources of data have been used in this study: US Census Bureau[30] and OpenStreetMap[38]. The geographic and population data are retrieved from the census at two levels: tracts and places. Census tracts are small, relatively permanent statistical subdivisions of a county which average 4,000 inhabitants. Places are defined as a concentration of population, housing, and commercial structures, identifiable by name. Specifically, the total population is associated to the variable B01003_001, taken from the 5-year American Community Survey which finished in 2021. Consequently, census tracts and places boundaries correspond to that year. These data can be accessed using the R package tidycensus[39]. For the urban road network, we used data from OpenStreetMap[38], queried from the Geofabrik website[40] in autumn 2023.

**OD matrices**

On one hand, a geographical intersection operation is carried out, utilizing boundaries on census tracts and cities. The centroids of all the census tracts that intersect or dwell inside the city limits were estimated. Only the centroids inside the city were selected as origins and destination points. At this stage, the Euclidean distance matrix between centroids has been calculated. On the other hand, road network data were supplied to a local server, downloaded from the OpenRouteService webapge[41]. This software considers how well certain streets are suited for the selected mode of transportation, i.e. car. Then, this priority is used in combination with the fastest option to yield the optimal route between centroids by using a contraction hierarchies' algorithm[42]. The connection between the main R program[43] and the server was made by an R API[44] which manages their communication. In this way, distances and times matrices for travels over the urban network have been obtained.

**Detour index**

The value of the detour index has been calculated in two ways: a linear regression and a statistical estimation. The first method was to plot the total travel distance between city centroids against the total Euclidean distance between the same points for all the set of cities. Obviously, the number of these routes for the k-city, $N_{R,k}$, is equal. Assuming a linear relationship between both distances, the proportionality constant between them for a given city, $C_k$, can be written as:

$$C_k = \frac{1}{N_{R,k}} \sum_{ij} D_{ij} \bigg/ \frac{1}{N_{R,k}} \sum_{ij} E_{ij} = \overline{D_{ij}}/\overline{E_{ij}} \quad (1),$$

where the bar denotes average. Therefore, $C_k$, the ratio between the total travel and Euclidean distances for the k-city, can also be understood as the mean ratio between the average travel length to the mean Euclidean distance. Equation (1) is referred in the literature as unweighted circuity[14]. However, by this method, the longest itineraries are the most favored. To avoid this, the second calculation method implies a double average, one over the routes of each city where each itinerary has the same contribution, $C_k = \overline{D_{ij}/E_{ij}}$, and the other over the previously calculated $C_k$ for the all the cities. It is important to note that while the first method allows us to identify cities with an anomalous detour index value, the second method enables us to examine the distribution of these indices in our set of cities.

**Population distribution fit**

Point pattern fitting typically involves proposing multiple models with several parameter values and comparing them with an experimental sample minimizing error estimation[45]. Our approach differs in that there is no real city perfectly fitting the Euclidean distance as presented by the Weibull function in Figure 4d. That is, we do not have an experimental point pattern to compare with.

However, we can propose a model of points that generates a Euclidean distance distribution as close as possible to our fit by minimizing the sum of squared errors between them. To achieve this, we begin with two fundamental principles: (a) centroids cannot overlap and (b) the probability of settling one centroid decays with distance from the center. The non-overlapping condition a) can be imposed using a hard sphere model with diameter $\phi$, while the attraction to the nucleus b) can be expressed as a decreasing probability distance, $(1-r)^\beta$. Here $\beta$ is the power of this dependency, and *r* varies between 0 and 1. Thus, a centroid is added when a) and b) are both satisfied simultaneously.

For the hard sphere diameter $\phi$ we have swept the range $[0,0.15]$ adopting an interval of 0.025. The attraction to the center is modulated by the expression $(1-r)^\beta$, where *β* belongs to the interval $[0,4]$ with an interval distance of 0.5. For each set of parameters within the model, 1,000 simulations were conducted, resulting that $\phi = 0.05$ and $\beta = 2$ generate the optimal fit. The number of centroids was set to 40, given that a greater population may result in a polycentric city case.

# References

1. Bettencourt, L. M. *Introduction to Urban Science: Evidence and Theory of Cities as Complex Systems* (John Wiley & Sons, Ltd, 2021).
2. Von Thünen, J. H. & Hall, P. G. *Isolated State* (Pergamon, 1966).
3. Krugman & Paul. Increasing Returns and Economic Geography. *J. Pol. Econ.* **99**, 483–499 (1991).
4. Alonso, W. *Location and Land Use: Toward a General Theory of Land Rent*. (Harvard University Press, 1965).
5. Frazer, J. The Reshaping of City Cores That Were Designed For Cars. *Forbes* https://www.forbes.com/sites/johnfrazer1/2019/08/06/the-reshaping-of-city-cores-that-were-designed-for-cars/ (2019).
6. Bettencourt, L. M. The origins of scaling in cities. *Science* **340**, 1438–1441 (2013).
7. Bettencourt, L. M., Lobo, J., Helbing, D., Kühnert, C. & West, G. B. Growth, innovation, scaling, and the pace of life in cities. *Proc. Natl. Acad. Sci. USA.* **104**, 7301–7306 (2007).
8. Succar, R. & Porfiri, M. Urban scaling of firearm violence, ownership and accessibility in the United States. *Nat. Cities.* **1**, 216–224; 10.1038/s44284-024-00034-8 (2024).
9. Lu, M., Zhou, C., Wang, C., Jackson, R. B. & Kempes, C. P. Worldwide scaling of waste generation in urban systems. *Nat. Cities.* **1**, 126–135; 10.1038/s44284-023-00021-5 (2024).
10. Molinero, C. & Thurner, S. How the geometry of cities determines urban scaling laws. *J. R. Soc. Interface* **18**, 20200705 (2021).
11. Boscoe, F. P., Henry, K. A. & Zdeb, M. S. A Nationwide Comparison of Driving Distance Versus Straight-Line Distance to Hospitals. *Prof. Geogr.* **64**, 188–196 (2012).

12. Chen, X. & Chen, Y. Quantifying the relationships between network distance and straight-line distance: applications in spatial bias correction. *Ann. GIS* **27**, 351–369 (2021).
13. Lee, M., Cheon, S. H., Son, S. W., Lee, M. J. & Lee, S. Exploring the relationship between the spatial distribution of roads and universal pattern of travel-route efficiency in urban road networks. *Chaos Soliton Fract.* **174**, 113770 (2023).
14. Giacomin, D. J. & Levinson, D. M. Road network circuity in metropolitan areas. *Environ. Plann. B Plann. Des.* **42**, 1040–1053 (2015).
15. Barthelemy, M. Spatial networks. *Phys. Rep.* **499**, 1–101 (2011).
16. Barthelemy, M. *The Structure and Dynamics of Cities: Urban Data Analysis and Theoretical Modeling* (Cambridge University Press, 2016).
17. Barthelemy, M. *Spatial Networks: A Complete Introduction: From Graph Theory and Statistical Physics to Real-World Applications* (Springer International Publishing, 2022).
18. Merchán, D., Winkenbach, M. & Snoeck, A. Quantifying the impact of urban road networks on the efficiency of local trips. *Transp. Res. Part A Policy Pract.* **135**, 38–62 (2020).
19. Barthélemy, M. & Flammini, A. Co-evolution of density and topology in a simple model of city formation. *Netw. Spat. Econ.* **9**, 401–425 (2009).
20. Zhao, F. X., Zhao, F. W. & Sun, H. A coevolution model of population distribution and road networks. *Physica A.* **536**, 120860 (2019).
21. Zhao, F., Wu, J., Sun, H., Gao, Z. & Liu, R. Population-driven Urban Road Evolution Dynamic Model. *Netw. Spat. Econ.* **16**, 997–1018 (2016).
22. Zhao, F., Sun, H., Wu, J., Gao, Z. & Liu, R. Analysis of road network pattern considering population distribution and central business district. *PLoS One* **11**, e0151676 (2016).
23. Louail, T. *et al.* Uncovering the spatial structure of mobility networks. *Nat. Commun.* **6**, 6007; 10.1038/ncomms7007 (2015).
24. Louail, T. *et al.* From mobile phone data to the spatial structure of cities. *Sci. Rep.* **4**, 1–12 (2014).
25. Xu, Y. *et al.* Urban dynamics through the lens of human mobility. *Nat. Comput. Sci.* **3**, 611–620; 10.1038/s43588-023-00484-5 (2023).
26. Mizzi, C. *et al.* Individual mobility deep insight using mobile phones data. *EPJ Data Sci.* **12**, 56 (2023).
27. Riascos, A. P. & Mateos, J. L. Networks and long-range mobility in cities: A study of more than one billion taxi trips in New York City. *Sci. Rep.* **10**, 4022 (2020).
28. Tenkanen, H. & Toivonen, T. Longitudinal spatial dataset on travel times and distances by different travel modes in Helsinki Region. *Sci. Data* **7**, 77 (2020).
29. Wang, P., Hunter, T., Bayen, A. M., Schechtner, K. & González, M. C. Understanding Road Usage Patterns in Urban Areas. *Sci. Rep.* **2**, 1001 (2012).
30. Unites States Census Bureau. https://www.census.gov/ (2023).
31. Gumbel, EJ. *Statistics of Extremes* (Columbia University Press, 1958).
32. Antal, T., Droz, M., Györgyi, G. & Rácz, Z. Roughness distributions for $1/f^{\alpha}$ signals. *Phys. Rev. E.* **65**, 046140 (2002).
33. Hansen, A. The Three Extreme Value Distributions: An Introductory Review. *Front. Phys.* **8**, 604053 (2020).

34. Lee, M., Cheon, S. H., Son, S. W., Lee, M. J. & Lee, S. Exploring the relationship between the spatial distribution of roads and universal pattern of travel-route efficiency in urban road networks. *Chaos Soliton Fract.* **174**, 113770 (2023).
35. Wang, F. & Guldmann, J. M. Simulating urban population density with a gravity-based model. *Socioecon. Plann. Sci.* **30**, 245–256 (1996).
36. Li, Y., Rybski, D. & Kropp, J. P. Singularity cities. *Environ. Plan. B: Urban Anal. City Sci.* **48**, 43–59 (2019).
37. Clark, C. Urban Population Densities. *J. R. Stat. Soc. Ser. A* **114**, 490 (1951).
38. OpenStreetMap. https://www.openstreetmap.org/ (2023).
39. Walker, K. & Herman, M. Load US Census Boundary and Attribute Data as 'tidyverse' and 'sf'-Ready Data Frames [R package tidycensus version 1.6.3]. Preprint at https://CRAN.R-project.org/package=tidycensus (2024).
40. Geofabrik. http://www.geofabrik.de/ (2023).
41. Openrouteservice. https://openrouteservice.org/ [Server version 7.1.0] (2023).
42. Geisberger, R., Sanders, P., Schultes, D. & Delling, D. Contraction Hierarchies: Faster and Simpler Hierarchical Routing in Road Networks. *Lecture Notes in Computer Science (including subseries Lecture Notes in Artificial Intelligence and Lecture Notes in Bioinformatics)* **5038 LNCS**, 319–333 (2008).
43. R Core Team & R Foundation for Statistical Computing. R: A language and environment for statistical computing. https://cran.r-project.org/ [R version 4.4.0] (2024).
44. Andrzej, O. Openrouteservice: Openrouteservice API Client. https://github.com/GIScience/openrouteservice-r/ [version 0.4.2] (2024)
45. Baddeley, A., Rubak, E. & Turner, R. *Spatial Point Patterns: Methodology and Applications with R* (CRC Press, 2016).

# Acknowledgements

The authors acknowledge funding from Spanish Ministry of Science and Education through project PID2021-123969NB-I00.

# Author contributions

D.O. and E.K. designed research; D.O. performed research; D.O. and E.K. wrote the manuscript. Both authors read, commented and approved the final version of the manuscript.

# Data availability

Original datasets can be downloaded from the census data webpage[30] and the Geofabrik server[40] (see the Methods section). Each OD matrix can be calculated via the supplementary code. Data on total lengths and times are also given via Github.

# Code availability

Scripts used for this study are available on the group's Github.

# Additional information

Competing financial interests: The authors declare no competing financial interests.

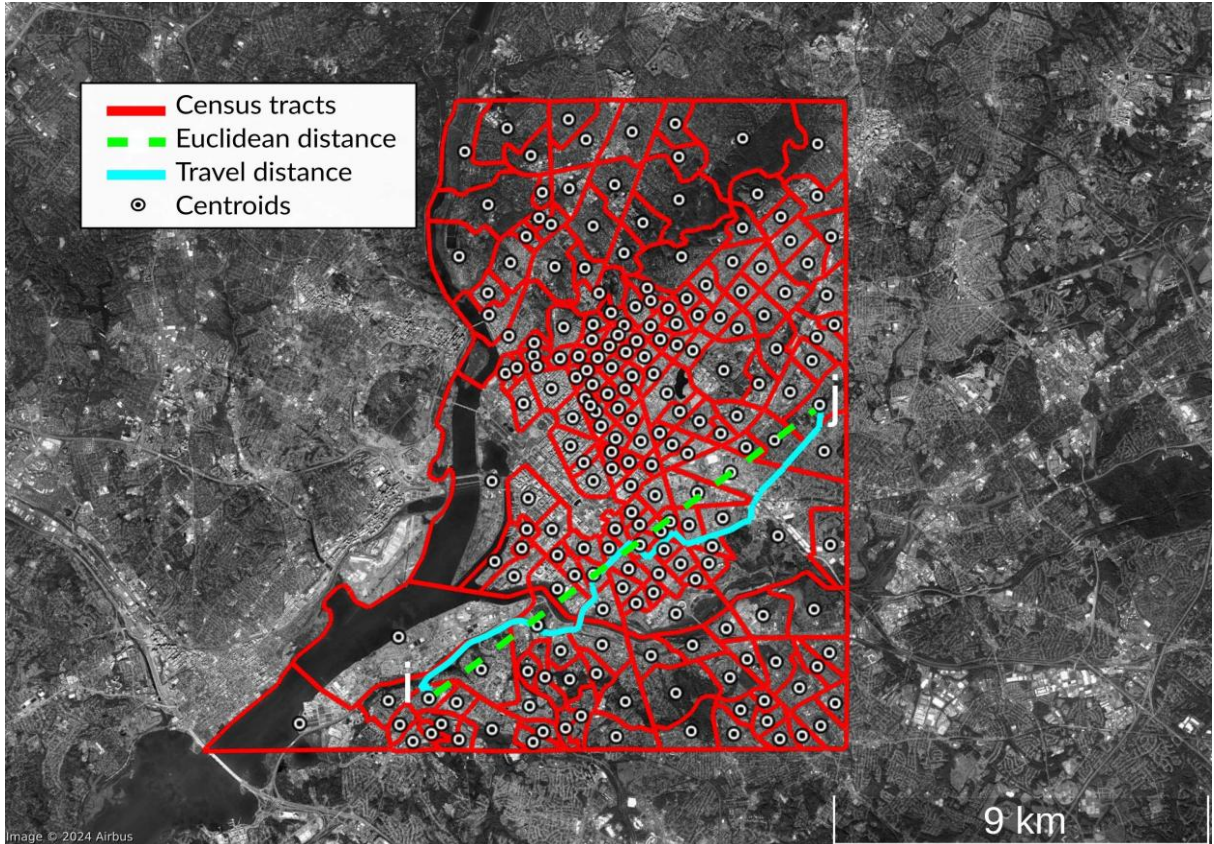

**Figure 1: Euclidean and travel distances between two centroids.** The city is divided by its census tracts, and the center of mass of each region is estimated. Taking as reference centroids i and j, the Euclidean distance is the length of the straight line that joins them. The travel distance is the length over the urban road network. These distances correspond to the matrix elements $E_{ij}$ and $D_{ij}$, respectively. Orthoimage of Washington D.C. courtesy of Google Earth.

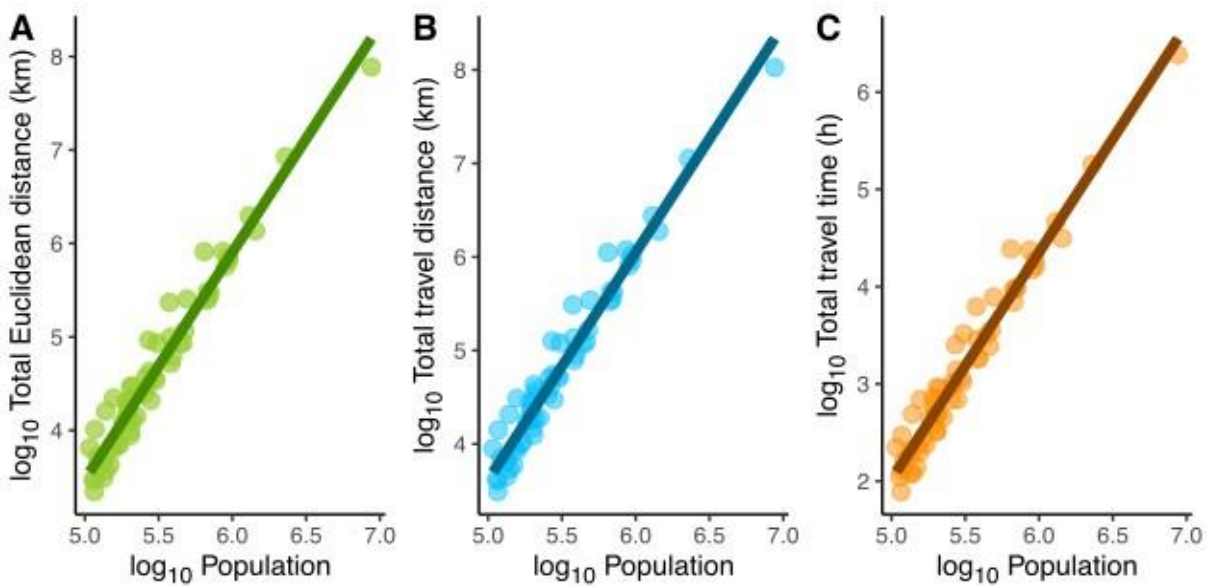

**Figure 2: Urban laws for routes in 60 North American cities.** Relation between the population, in the x-axis, versus **a)** the total length of the Euclidean distances connecting centroids, **b)** the total length of travel distances and the **c)** total time of travels.

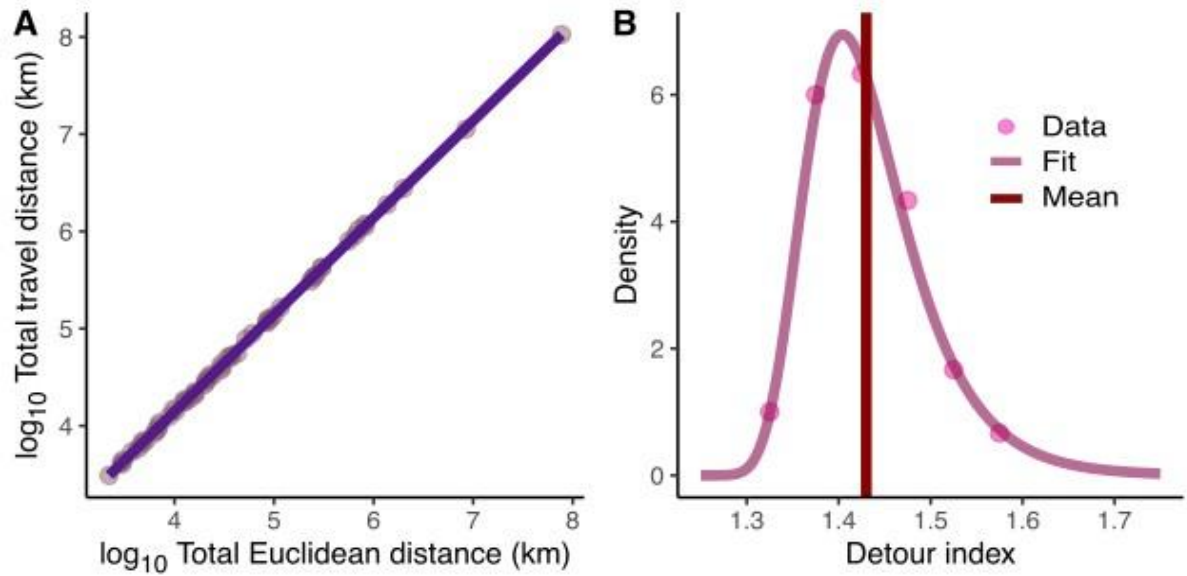

**Figure 3: Mean detour index estimations. a)** Total length of the travel versus its corresponding Euclidean distance. Average detour index is calculated by mean its slope. **b)** Density histogram of the detour index calculated as a mean for every city and fitted by a Gumbel function. The mean detour index is averaged from the previous means.

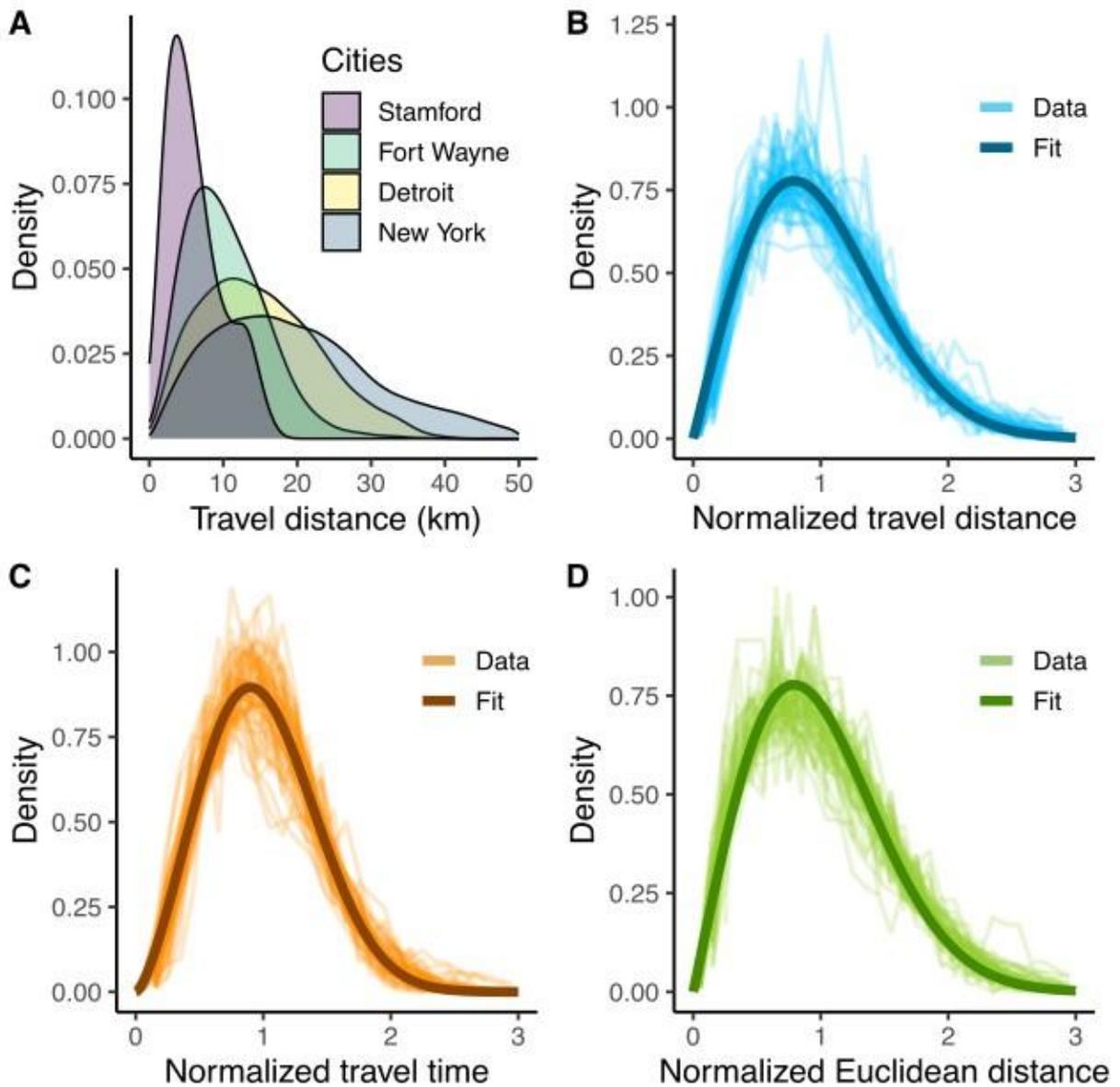

**Figure 4: Collapse of travel and distance distributions. a)** Density histogram of travel distances for various cities, ordered by population. Normalized density histograms using each city mean for **b)** travel distances **c)** travel time and **d)** Euclidean distances between centroids. The fitted lines are given by Weibull functions.

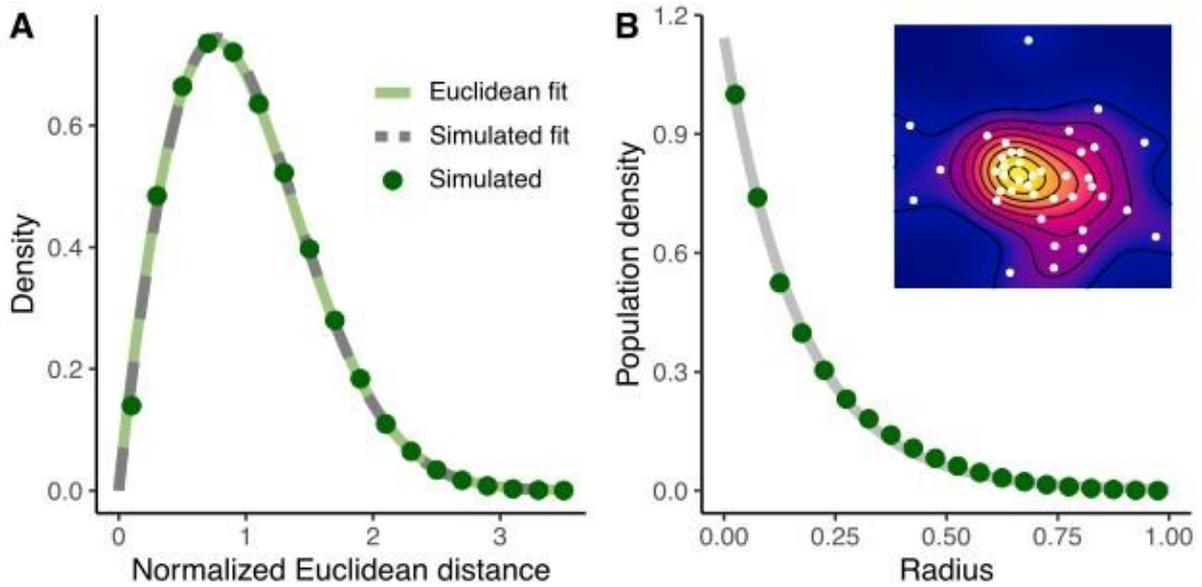

**Figure 5: Simulated distributions of distances and population for 1,000 cities. a)** Normalized density histogram of simulated Euclidean distances between centroids compared to the Weibull curve from Figure 4b. The latter is obtained as a fit to the set of real cities. **b)** Radial population density fitted by a decreasing exponential function. The inset shows one run of the model, with the centroids represented by white dots and highest density zones colored in brighter tones.